# Leveraging LLMs and Social Media to Understand User Perception of Smartphone-Based Earthquake Early Warnings

Hanjing Wang[1], S. Mostafa Mousavi[1,2*], Patrick Robertson[4], Richard M. Allen[2,3], Alexie Barski[2], Robert Bosch[2], Nivetha Thiruverahan[2], Youngmin Cho[2], Tajinder Gadh[2], Steve Malkos[2], Boone Spooner[2], Greg Wimpey[2], Marc Stogaitis[2]

[1] Department of Earth and Planetary Sciences, Harvard University, Cambridge, MA, USA.
[2] Google LLC, Mountain View, CA, USA.
[3] Seismological Laboratory, University of California, Berkeley, Berkeley, CA, USA.
[4] Google Germany GmbH, Munich, Germany.
*Corresponding author. Email: mousavim@google.com

## Abstract

Android's Earthquake Alert (AEA) system provided timely early warnings to millions during the Mw 6.2 Marmara Ereğlisi, Türkiye earthquake on April 23, 2025. This event, the largest in the region in 25 years, served as a critical real-world test for smartphone-based Earthquake Early Warning (EEW) systems. The AEA system successfully delivered alerts to users with high precision, offering over a minute of warning before the strongest shaking reached urban areas. This study leveraged Large Language Models (LLMs) to analyze more than 500 public social media posts from the X platform, extracting 42 distinct attributes related to user experience and behavior. Statistical analyses revealed significant relationships, notably a strong correlation between user trust and alert timeliness. Our results indicate a distinction between engineering and the user-centric definition of system accuracy. We found that timeliness is accuracy in the user's mind. Overall, this study provides actionable insights for optimizing alert design, public education campaigns, and future behavioral research to improve the effectiveness of such systems in seismically active regions.

# Introduction

EEW systems offer crucial seconds to minutes of warning before damaging ground shaking reaches urban areas (Nakamura, 1988; Allen et al., 2009). These systems work by rapidly detecting the initial, less destructive seismic waves near the earthquake's origin. They then quickly assess the potential hazard and issue alerts to affected regions (Heaton, 1985). The effectiveness of EEW systems rely on electronic signals traveling faster than seismic waves, creating a vital window for protective actions. Such measures include seeking cover or halting industrial operations which can significantly reduce casualties and economic damage during an earthquake (Allen & Melgar, 2019).

In response to the substantial cost of infrastructure and maintenance for traditional EEW systems, smartphone-based EEW systems have emerged, utilizing the low-cost accelerometers found in modern mobile devices to create distributed seismic sensing networks (Kong et al., 2016; Finazzi, 2016; Minson et al., 2015; Finazzi et al., 2024). This approach offers a valuable alternative or complement, potentially extending coverage to areas with sparse sensor networks or enhancing existing infrastructure. A notable example is Google's AEA system, which leverages billions of Android smartphones to form a global seismic network, capable of real-time earthquake signal detection and near-light-speed alert transmission via the internet (Google, 2020, Allen et al., 2025).

On April 23, 2025, a Mw 6.2 earthquake struck the central Marmara Sea, Türkiye, at 12:49 local time (09:49 UTC). Occurring at a depth of 10 km (USGS, 2025; GFZ, 2025), this event was the largest in the region in 25 years (Hubbard and Bradley, 2025), the last of a long sequence of large earthquakes on the Northern Anatolian fault (Şengör and Yilmaz 1981; Şengör et al., 2005). . Initial reports indicated that buildings in Istanbul, 75 km northeast of the epicenter, experienced moderate shaking at MMI V. At least 359 people in the city were injured, primarily due to panic rather than direct structural damage (Tagesschau, 2025; Memişoğlu, 2025) (Figure 1).

The Mw 6.2 event served as a crucial real-world case study for evaluating Google's AEA system. The system successfully delivered alerts to millions of users with high precision providing over a minute of warning before the strongest shaking (Mousavi et al., 2026). To assess the interaction of the users with AEA's alerts, public social media posts were analyzed in this study. LLMs were used for automated data processing and information extraction from user posts on the X platform. Rather than a purely exploratory analysis, this study evaluates the user experience by testing a core hypothesis derived from existing risk communication literature: A user's perceived accuracy and the informational clarity of an alert significantly dictate their perception of the system's usefulness and their future trust in it. Testing these hypotheses offers actionable insights into the practical role of smartphone-based EEW in mitigating seismic risk.

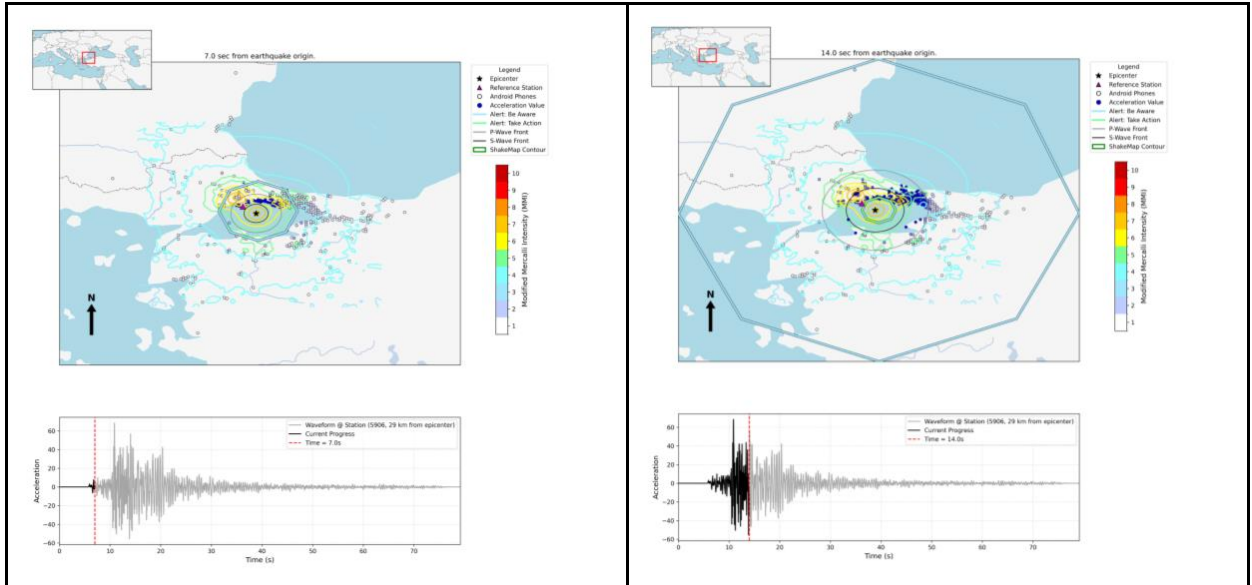

Figure 1: The first AEA contour, issued ~ 7 s after the earthquake origin time, (left) and a subsequent alert issued 7 seconds later (14 s after the earthquake origin time) (right) are shown by polygons overlaid on top of USGS ShakeMap contours color coded based on the intensity of shaking. Normalized median acceleration values of all phones within 4x4km grids are shown by filled circles (in blue). The waveform on the bottom, is from a reference station (shown by purple triangle on the map) from the Disaster and Emergency Management Authority (AFAD) seismic network. The earthquake was detected by Android phones shortly after the earthquake waves (P wave) hit the shore and an AEA alert was sent preceding the earthquake S waves at most of the locations. The alert polygon was rapidly expanded to a large area as the magnitude estimate grew while the shaking was still ongoing in the epicenter's surrounding area (right).

## Data

To analyze user experience, we collected 511 public posts on the social media platform X over a 3 days period following the Mw 6.2 Marmara earthquake. This specific time frame was chosen to capture immediate, organic user reactions and subsequent reflections while mitigating the risk of memory degradation that occurs over longer periods. These posts are specifically relevant to AEA alerts issued during this time and served as our data source for evaluating the user experience. We collected relevant posts by employing a single keyword, "Android Deprem," ("deprem" means "earthquake" in Turkish) and as screenshots that contained both text and images. This targeted keyword was intentionally adopted to prioritize precision over recall. By focusing on a specific, localized keyword, we filtered out the overwhelming noise of global news bots and international observers, ensuring the dataset consisted primarily of actual alert recipients in the affected region. While this localized search may have omitted some posts written entirely in other languages without the local keyword, it successfully yielded a concentrated, high-quality sample for deep behavioral extraction. Images included in the posts often are screenshots of user-received alerts

(Figure 2). These alert screenshots were displaying key information such as the alert timestamp, the estimated earthquake magnitude at the time of issuance, the alert's geographical contour, and the user's approximate location within the alerted zone. This data, independent of the AEA, is collected passively unlike other user experience surveys. Its comment-based and free-form nature offers the flexibility to explore a broader spectrum of user experiences and behaviors during the earthquake. By leveraging unstructured social media data from the X platform and the capabilities of LLMs, we were able to extract detailed information regarding alert types (Take Action vs. Be Aware), alert mode (silent or audible), and subsequent post-alert actions. The dataset also captured user sentiment and specific emotions, along with recommendations for system improvements.

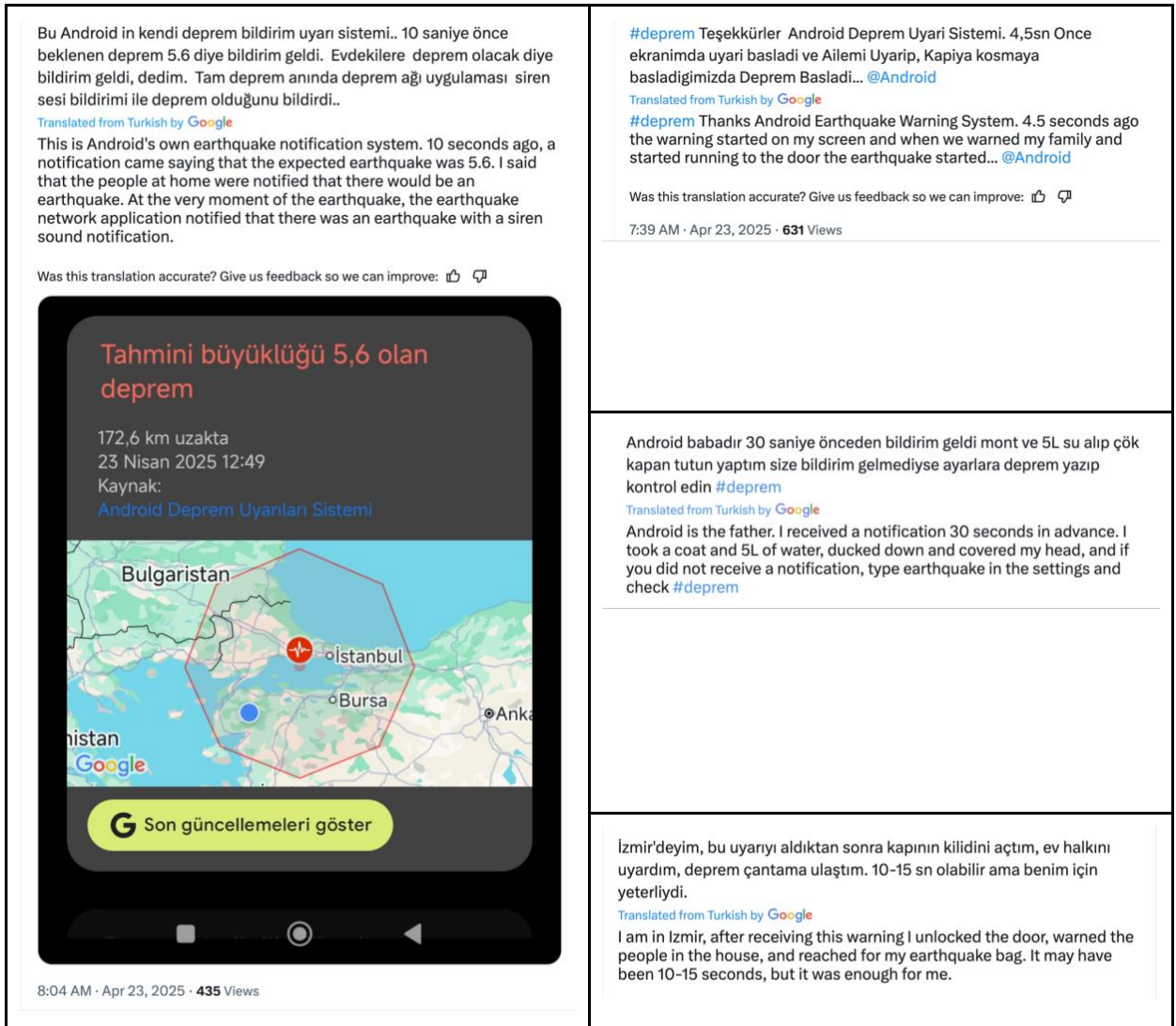

Figure 2: Four representative X posts used in this study.

Furthermore, posts encompassed perceived shaking intensity, whether the alert was the user's first EEW experience, user gender, and recalled alert information. The data also included perceived accuracy and clarity of alert information, technical issues encountered, reasons for inaction post-alert, and rationales for alert helpfulness. Finally, it included a comparative assessment of alert delivery and precision relative to other EEW alerts received for the same event, as well as reported warning times. The diverse attributes extracted from X posts offer a chance to examine further aspects of alerting effectiveness and the complex interplay among these factors. However, relying on a singular social media platform (X) and a single keyword ('Android Deprem') introduces inherent sampling bias into the collected dataset, capturing only a self-selected fraction of users who chose to post publicly. Furthermore, the presence of another operational Android-based EEW application in the region during the earthquake might contribute additional noise to the data. Therefore, all collected posts first underwent a visual inspection for data cleaning, followed by an automated pre-processing step, to make sure only AEA related posts (the resulting 511 posts) are used for the subsequent data analyses.

# Methods

To effectively process the social media posts from the X platform and extract pertinent information, LLMs were employed. Specifically, Google's Gemini Pro 2.5 (Gemini Team, 2023) and Vertex AI, Google Cloud's comprehensive machine learning platform, were utilized to extract 42 distinct attributes (encompassing both questions and demographic data) from these posts through careful prompt engineering and few-shot learning methodologies (Figure 3). Few-shot learning is a machine learning paradigm where models are trained on a limited number of examples (shots) for a new task, enabling rapid adaptation and generalization without extensive retraining. This approach leverages pre-trained knowledge from a broader dataset, allowing the model to quickly learn and perform new tasks with minimal task-specific data. The considerable potential of LLMs in computational social science is underscored by their capacity for advanced analysis of unstructured data, leading to the derivation of actionable insights. This capability is reflected in their escalating application within natural disaster management, earthquake science, and studies of user behavior (e.g., Mousavi et al., 2025; Lei et al., 2025; Xu et al., 2025; Linardos et al., 2025; Raj et al., 2025; Zhou et al., 2025). The queries incorporated into the prompt were specifically designed to assess user attitudes, preferences, and actions in response to AEA alerts, while also integrating additional information for data validation and refinement. Details regarding the designed prompt, the extracted information, and subsequent data visualization are provided in the supplementary materials. To ensure the accuracy of the attributes extracted from the X posts, the distributions of various auxiliary attributes were cross-referenced for consistency, and any entries exhibiting suspicious attribute values were excluded. Furthermore, a random subset comprising 10% of the data for all extracted attributes underwent manual auditing to confirm the precision of the LLM outputs. This rigorous manual validation corroborated the reliability of the extracted information.

Emotional reactions, including annoyance, reassurance, excitement, surprise, fear, and anxiety, were determined through the automated analysis of tweet content using LLMs. Gemini was trained with careful prompt engineering and few-shot learning methodologies to identify and categorize these specific emotions expressed by users in their social media posts. This allowed for the extraction of nuanced emotional states beyond general sentiment, providing deeper insight into user experience. To provide concrete evidence of Gemini's effectiveness in sentiment analysis under disaster scenarios, we manually validated the model's emotional estimates against the original Turkish posts. This manual cross-validation confirmed that Gemini accurately captured nuanced emotional states—even amidst the chaotic, informal syntax typical of disaster-related social media.

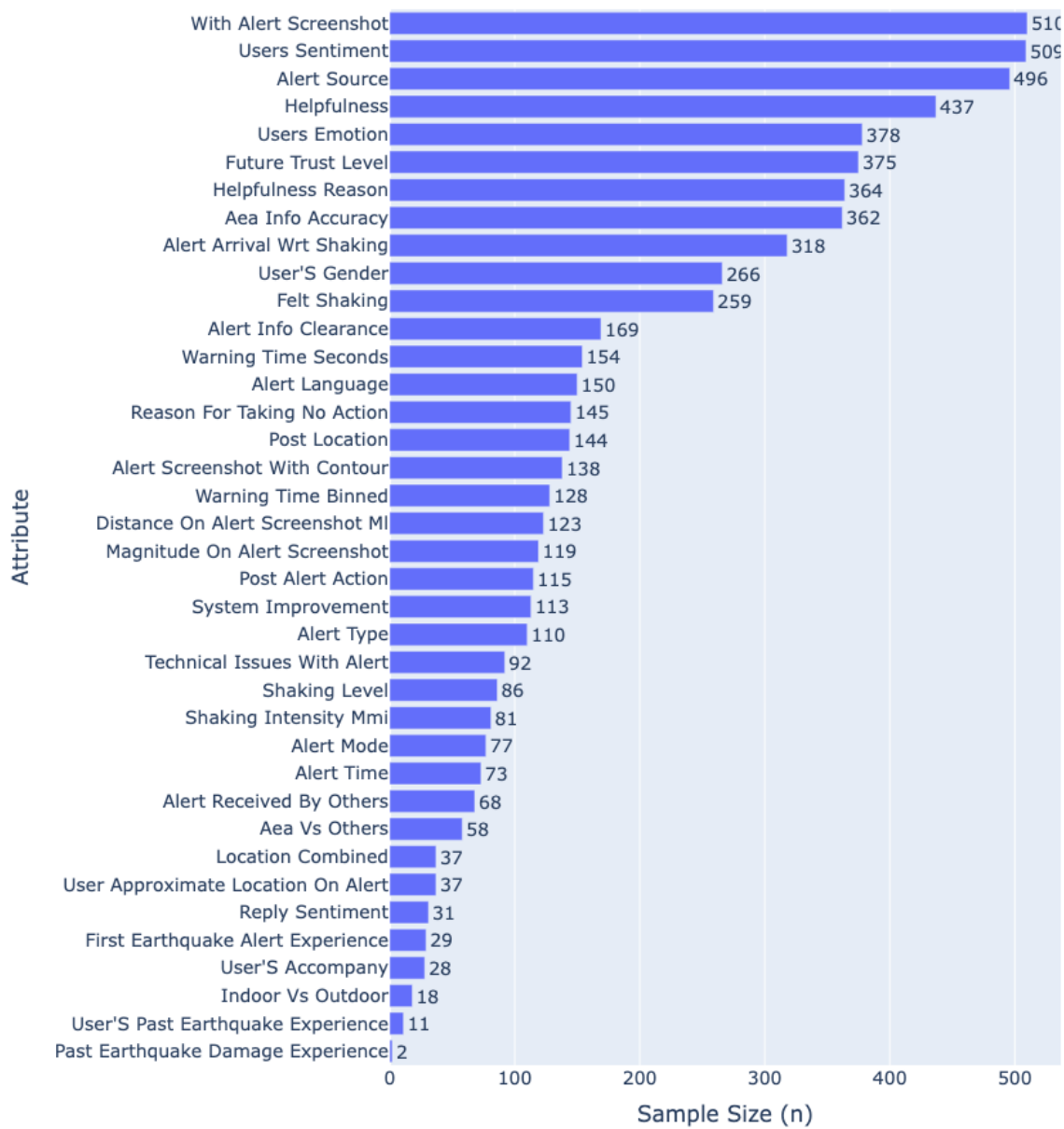

Figure 3: Distribution of attributes extracted from X posts. There are much fewer posts with information regarding user's past earthquake experience as compared to the information regarding users' emotion, trust level, etc. "location combined" feature represents aggregated geolocations into local metropolitan areas.

To ensure the validity of extracted information by Gemini, we have manually cross-checked Gemini's output with the input posts for multiple key attributes. Upon cross-referencing all posts (26) for the type of Alert, it was found that almost all were correctly classified, with the exception of one "TAKE_ACTION" classification that was actually a "BE_AWARE" alert. Cross-checking 44 posts for Alert mode classifications revealed that most ALERT_WITH_SOUND classifications were accurate, with only about five instances appearing to be screen notifications without sound, which are less likely to prompt action. Conversely, many SILENT_NOTIFICATION classifications seemed to indicate no notification at all, as only one of the seven checked posts actually received a silent notification after the earthquake. Evaluating inferred alert arrival (all posts cross-checked) showed that nearly all extracted alert arrival times were accurate. Most "AFTER_SHAKING" classifications corresponded to alerts that never arrived, while a few "DURING_SHAKING" alerts appeared to be slightly before the shaking arrived.

The cross-validation of the inferred emotional accuracy (across all samples) indicated that the majority of the extracted perceptions were reliable. Further internal evidence of the LLM's accuracy is demonstrated by the logical consistency of the extracted sentiments with subsequent user behavior. Although some identified actions were slightly inaccurate (e.g., certain "drop-and-cover" instances derived from screenshot advice, where users were only passively aware), Gemini generally demonstrated the ability to differentiate between a lack of action and various levels of passive or active awareness. An examination of the original posts regarding the emotions inferred by Gemini reveals a strong correlation between positive sentiments, specifically reassurance and gratitude, and the inclination to take action, such as evacuation or engaging in drop-cover maneuvers. Conversely, instances of annoyance primarily stemmed from issues with Android alert delivery, either delays in notification or complete failures to send alerts after the earthquake. A portion of this annoyance was also directed at iPhones, with users expressing a desire to switch to Android due to perceived inconsistencies in alert functionality.

# Results

## Alerting Performance

It is challenging to extract precise geolocation data from X posts. User location estimates are based on information within their posts or attached alert screenshots. Data analysis shows that over 56.7% of users in the Istanbul metropolitan area received "Be Aware" (BA) alerts, while 13.4% received "Take Action" (TA) alerts (Figure 4). This aligns with Istanbul experiencing moderate shaking, estimated at MMI V[1]. Based on our extracted data, of the posts where alert arrival relative to shaking could be determined (n=78), 55% reported receiving the alert before the shaking began, compared to 36% after and 9% during. Posts from other locations exclusively contained BA alerts, which aligns with the extent of alert contours (Figure 1). In Istanbul, alert levels generally aligned with the perceived shaking intensity. Specifically, 49% of reports indicated weak shaking, while 8.2% reported strong shaking. The estimated ground shaking intensity across the affected area ranged from MMI-III to MMI-VII, with more frequent reports of MMI-III shaking. This data is consistent with USGS reports (Figure 4, top panel).

Consistent with these arrival times, the majority of posts reported a warning time ranging from 0 to 60 seconds (Figure 5b). This pre-shaking warning time is crucial for users to take protective actions. The user sample analyzed in this study primarily consisted of individuals located within 300 km from the earthquake's epicenter (Figure 5c). There were 35 posts (~ 7%) with negative sentiments directly related to the AEA, among which 4 users stated that the provided warning time was not enough to take an active action. The rest of the posts were related to users who did not receive AEA alerts while expecting them.

The majority of X users mentioned feelings of the earthquake shaking similar to the results of the AEA's in-alert user surveys (Figure 6), However, the percentage of "Not Felt" responses in AEA's survey results is higher than those in the X posts. This could be due to a fact that AEA surveys cover a bigger geographic area extended to the lower intensity regions at the margin of the affected area while the majority of the X posts come from users in Istanbul metropolitan (Figure 4). Approximately 20% of the users reported the receiving of the alert after feeling the shaking in both datasets, However, the percentage of the X users that reported the receiving of the alert during the shaking is less that what has been reported in AEA survey. This can be an indication of a bias in X posts.

These hierarchical distributions (Figure 7) illustrate the complex interplay between technical performance and user psychology. The data shows that "Be Aware" alerts were most frequently received before the onset of shaking. Critically, the hierarchy demonstrates that alert arrival time relative to shaking is a primary determinant of "future trust," reinforcing the study's conclusion that receiving a warning before shaking is the most powerful factor in building system credibility.

Figure 8 categorizes the human element of the disaster, showing the distribution of post-alert actions and emotional states. The emotional distribution highlights that while fear and anxiety were present, more positive sentiments like reassurance and surprise were the dominant among the social media users. Although it would be interesting to study the relation between the user's state of emotion and their post alert actions, the low number of samples prevents us from performing more detailed statistical analyses in this study.

Figure 4: a) Alert type by location, b) alert arrival relative to shaking, c) shaking level, and d) user sentiments extracted from X posts categorized by location.

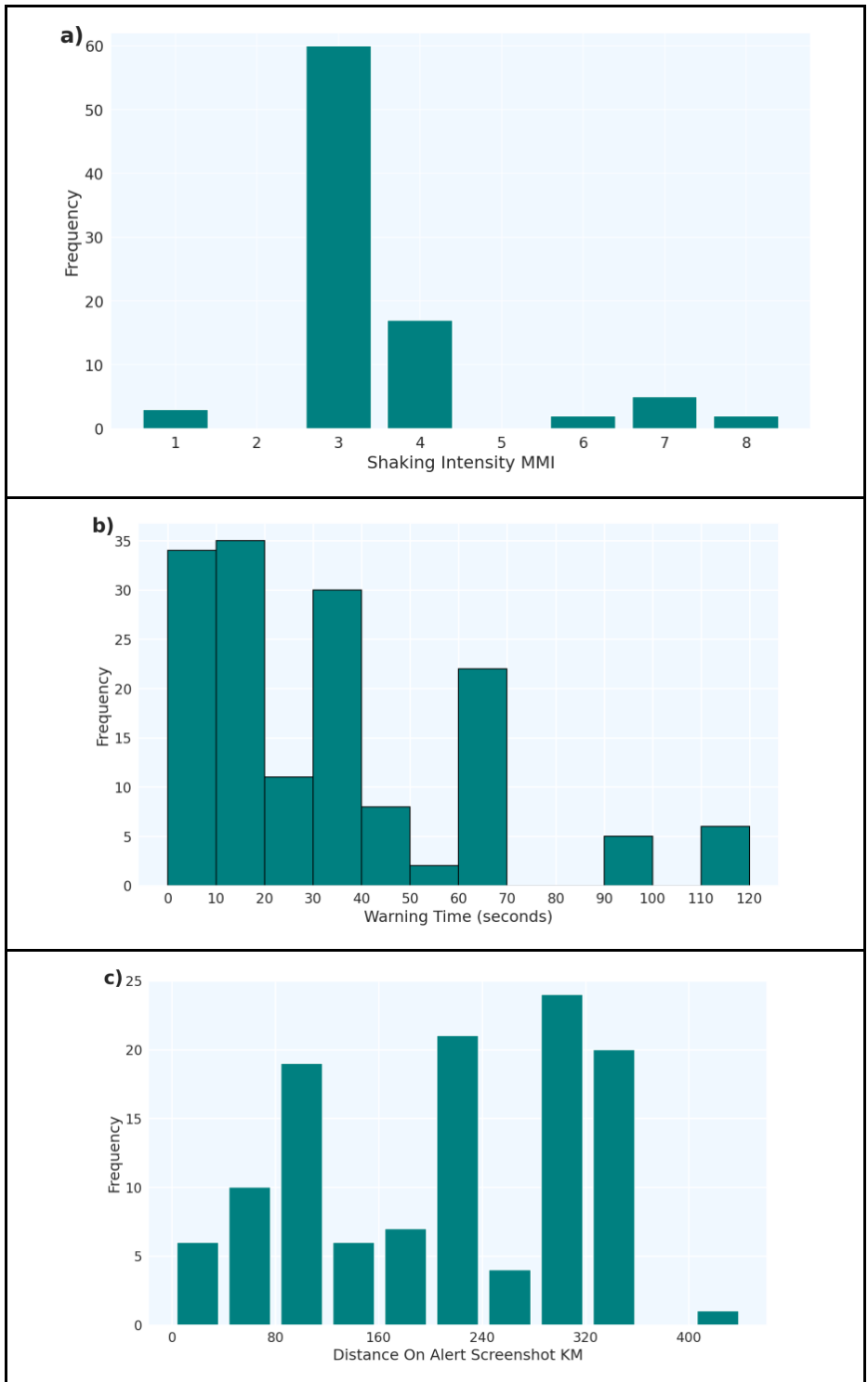

Figure 5: Estimated ground shaking intensity (a, n=81), received warning time (b, n=154), and epicentral distance in kilometers (c, n=123) from X posts. Note that because data is extracted from unstructured social media posts, not all posts contain information for every attribute; therefore, the sample sizes (n) differ across the subpanels and do not represent perfectly overlapping sets of users.

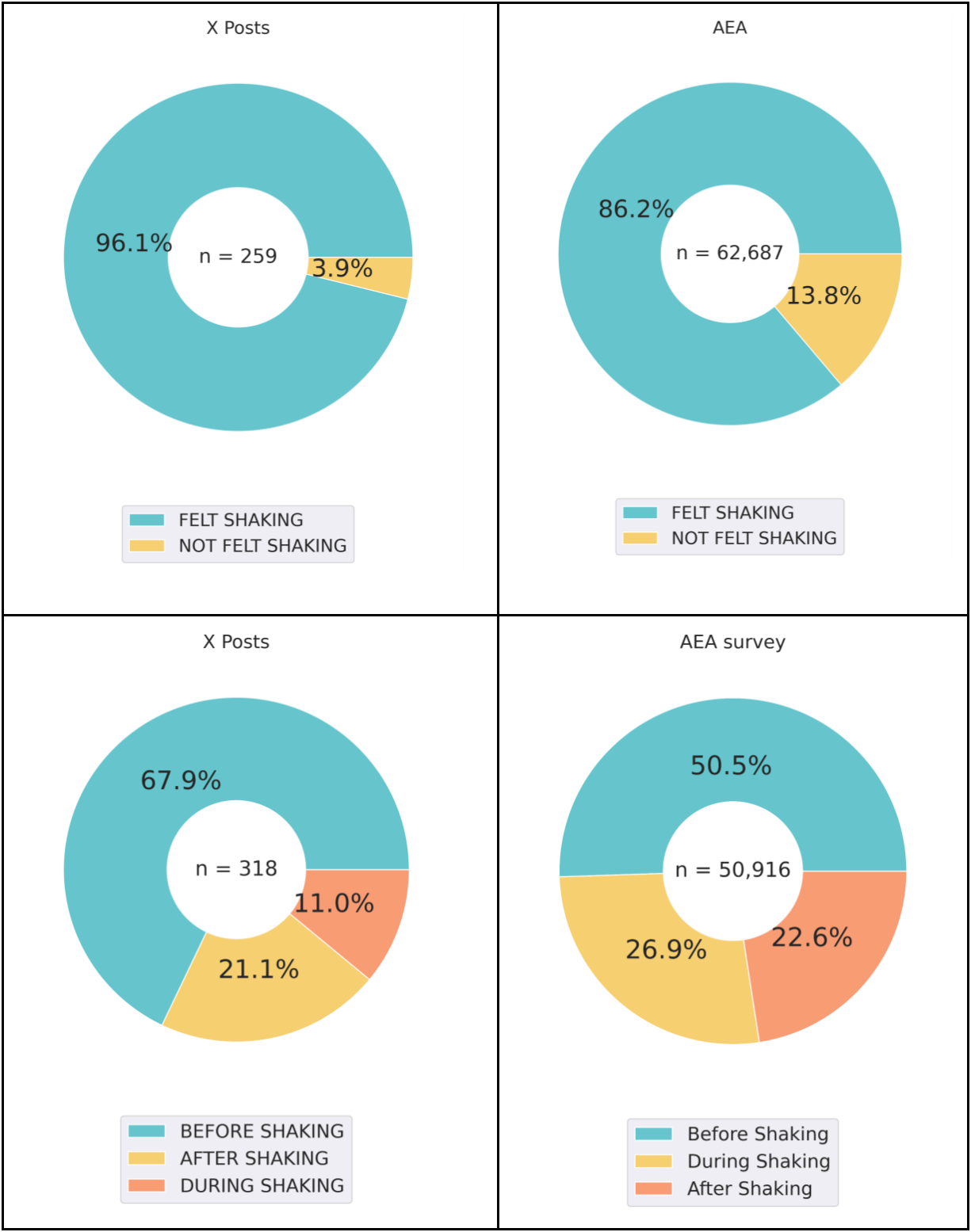

Figure 6: Distribution of responses to "When did you receive the alert?" (top) and "Did you feel shaking?" (bottom) questions in X posts (left) and AEA feedback survey (right).

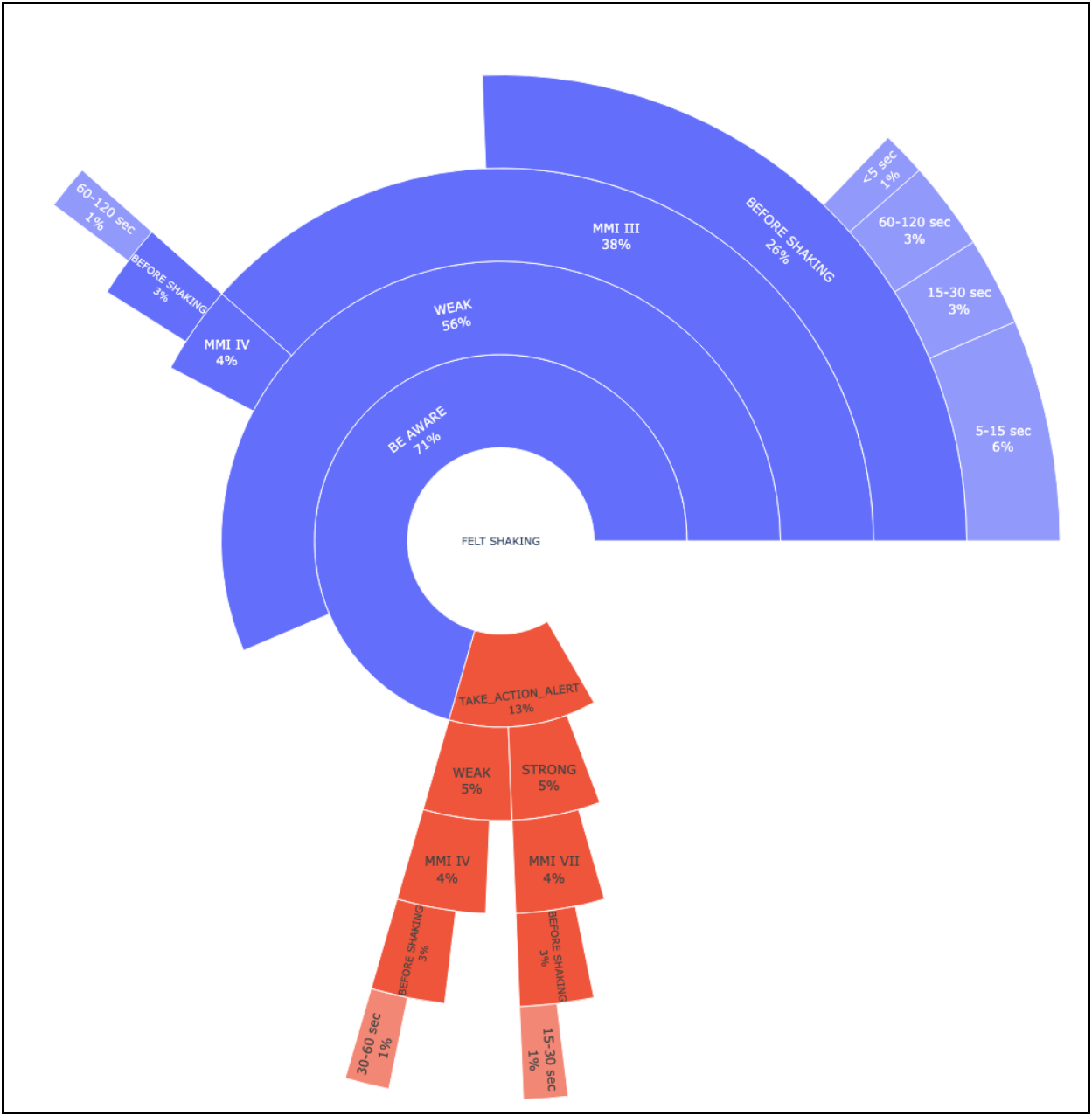

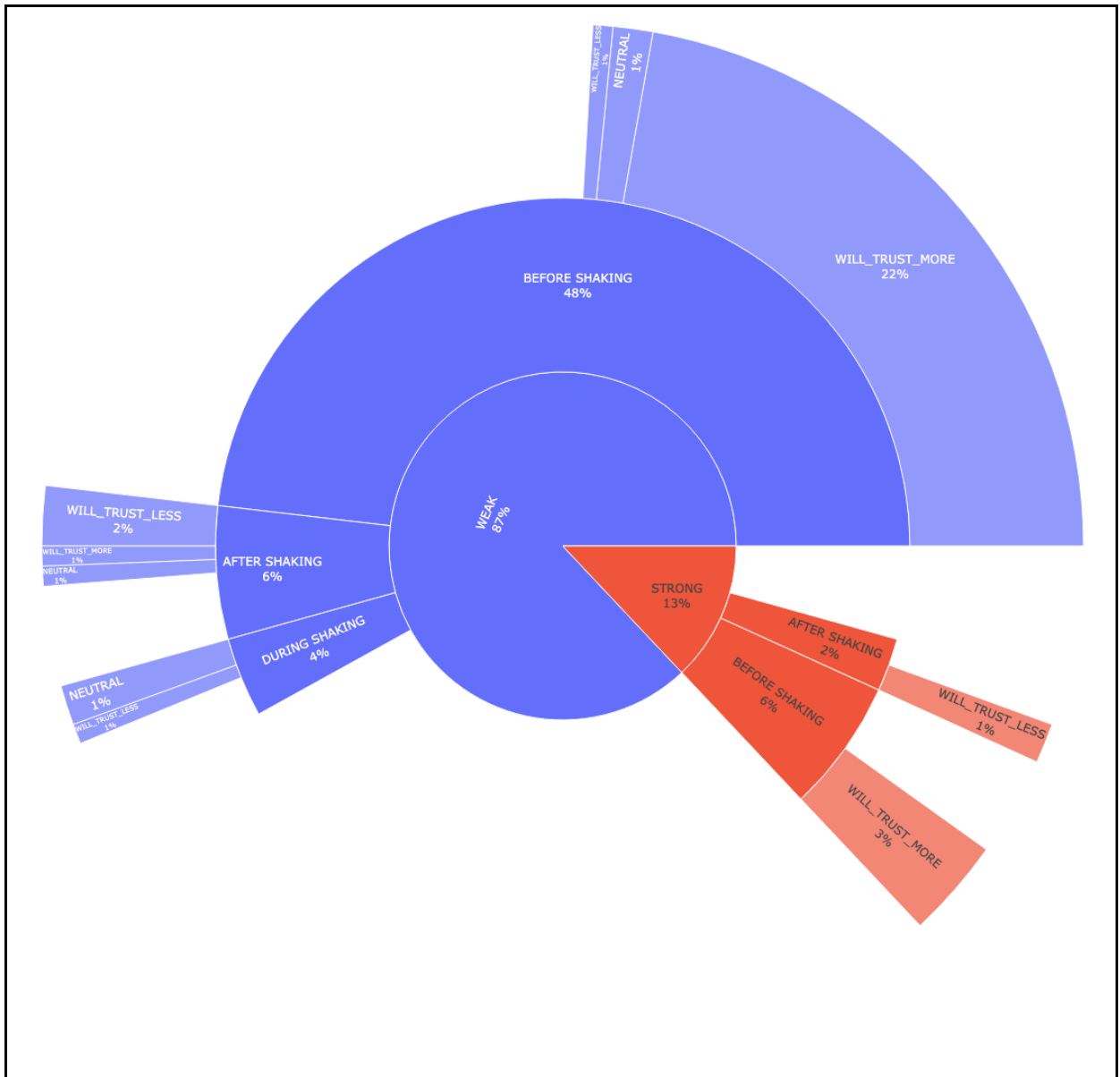

Figure 7: Two examples of hierarchical distributions of several key attributes extracted from X posts. Specifically, it illustrates: top) felt shaking, alert type, shaking level, estimated intensity, alert arrival relative to shaking, and warning time; bottom) shaking level, alert arrival relative to shaking, and future trust (n=30).

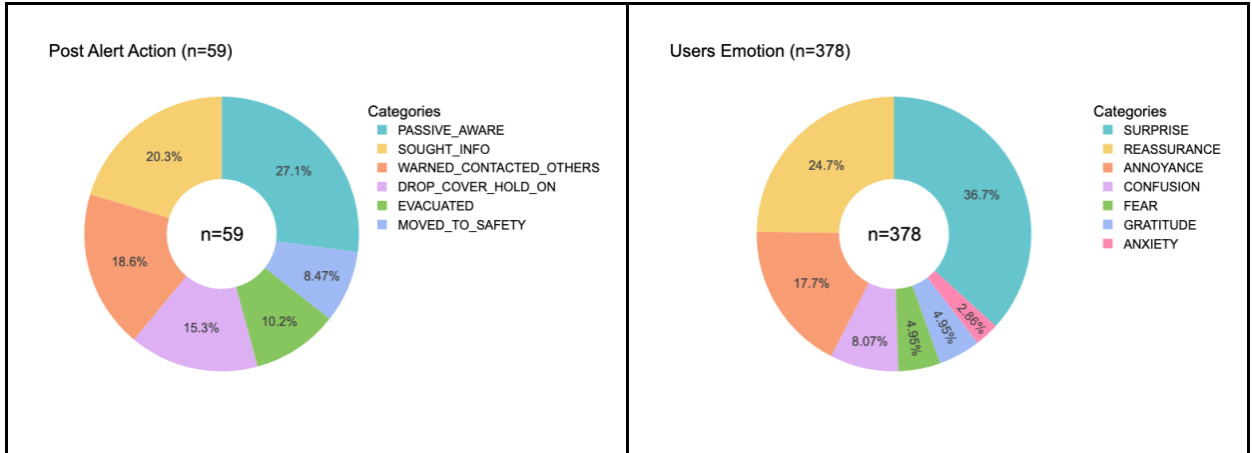

Figure 8: Distribution of the post alert actions (left) and emotions (right) among X users.

# Variable relationships

Following the methodology of Goltz et al. (2024), we apply nonparametric statistical methods to analyze social media datasets, evaluating statistical significance with conventional thresholds ($p<0.05$, $p<0.01$, and $p<0.001$). The Pearson chi-square test for independence was used to examine the relationship between two categorical variables. This test compares observed values to expected values, assuming no association between the variables.

### Reported usefulness of the alert

A user's perception of an alert's accuracy is the most important factor in determining its usefulness. Precise alerts (from user's perspective) are consistently rated as useful, while inaccurate ones are overwhelmingly dismissed as unhelpful. This raises a crucial question for the AEA System, especially for regions like Turkey where the 2025 M6.2 earthquake may have been many users' first experience with both a major earthquake and the alert system. Without prior exposure, how can we ascertain if users perceived these alerts as precise? Does a pre-existing trust in Android/Google systems inherently lead to greater alert credibility, even for novice users? This critical aspect requires further exploration as user trust and perceived accuracy are important to the system's effectiveness. Our analysis shows a user's neutral sentiment remains independent of the alert's accuracy (Table 2).

TABLE 2: Statistically Significant Variables Influencing Alert Usefulness.

| Variable | Chi-square Statistic | DF | Significance | Relationship | Total observations (N) |
|---|---|---|---|---|---|
| Perception of alert | 279.11 | 2 | $p < 0.001$ | If a user perceives an alert as inaccurate, they are overwhelmingly | 344 |

| | | | | likely to also find it not useful. | |
|---|---|---|---|---|---|
| information accuracy | | | | | |
| Emotional reaction | 184.39 | 10 | p < 0.001 | The emotional state of a user is strongly associated with perceived alert's usefulness. | 348 |
| Alert arrival wrt shaking | 175.02 | 6 | p < 0.001 | There is a very strong association between receiving an alert before shaking and perceiving it as useful. | 286 |
| Alert information clearance | 39.14 | 2 | p < 0.001 | A "Useful" rating is almost exclusively associated with clear information. | 160 |
| Duration of warning time | 26.70 | 8 | p < 0.001 | As the warning time increases, the perception of the alert shifts away from "Neutral" and towards "Useful". | 112 |
| Alert mode | 18.84 | 2 | p < 0.001 | The absence of sound is a strong predictor that a user will find the alert not useful. | 69 |

Regardless of whether users felt "reassured" or "annoyed", they overwhelmingly found the alert useful; neither of these emotional responses correlated with a "Not Useful" rating. Conversely, a strong association was observed between "confusion" and negative or neutral perceptions of the alert's utility. "Fear" and "surprise" were also associated with less positive assessments, leading users toward neutral or negative evaluations.

The timeliness of an AEA alert significantly influenced its perceived utility. In other words, timeliness is accuracy in the user's mind. Alerts received before ground shaking were strongly correlated with a positive assessment of their usefulness, whereas alerts received during or after shaking were overwhelmingly considered unhelpful. Specifically, among those who received the alert before shaking began, 28% rated it as somewhat or extremely useful. Conversely, alerts received during or after ground shaking had a significantly lower positive reception, with only 1.0% and 1.5% rated positively. Negative evaluations of usefulness were considerably higher at 3% and 8%.

The clarity of information presented within an alert strongly associated with its perceived utility. Our findings indicate that for an alert to be deemed useful, its content must be clear. Conversely, unclear information significantly increases the likelihood that users will rate the alert as "Neutral" or "Not Useful." Furthermore, analysis of the relationship between alert clarity and reported helpfulness demonstrates a strong connection to timeliness and practical utility. Clear alerts are consistently seen as useful, whether they offer confirmation or vital preparation time. Nonetheless, the tumultuous conditions during ground shaking frequently lead to unclear alerts, hindering comprehension. Conversely, alerts issued prior to the onset of shaking are more readily understood (Table S1). The perceived usefulness of an alert was also statistically significantly associated with the duration of warning time. An increased warning time led to a shift in the perception of the alert from "Neutral" to "Useful." Specifically, warning times exceeding 30 seconds were strongly associated with a higher utility rating for the alert. A "Neutral" assessment of usefulness was more frequently associated with very short warning times (under 15 seconds).

The alert's auditory component critically influences its perceived utility. A significant correlation exists between silent alerts and those rated "Not Useful," while audible alerts (not including vibrate mode) were less prone to negative feedback.

## User's future trust

User trust in an alert system is often a direct reflection of their past experiences. However, trust isn't always built on repeated individual interactions; it can also stem from an implicit belief in the system's provider (e.g., Google/Android's reputation for useful technology) or from public education by government authorities regarding the earthquake risks and preparations. Therefore, while trust is typically earned through multiple accurate alerts, it is plausible for a single critical event to establish trust in a system that delivers vital seismic information, especially in regions with a history of significant earthquake damage and fatalities.

Our analyses indicate that user confidence and trust in a system are primarily built on helpful alerts, while unhelpful experiences significantly erode that trust. Cultivating user trust heavily depends on providing actionable preparation time. Conversely, an untimely alert not only fails to build trust but actively diminishes it.

TABLE 3: Variables Statistically Significant to a User's Future Trust.

| Variable | Chi-square Statistic | DF | Significance | Relationship | Total observations (N) |
|---|---|---|---|---|---|
| Helpfulness | 664.61 | 6 | p < 0.001 | A helpful alert is the single most important factor in building a user's confidence and trust in the system. | 374 |
| Helpfulness reason | 334.94 | 4 | p < 0.001 | Providing actionable prep time is the most powerful way to build user trust. | 339 |
| Alert information accuracy | 282.79 | 2 | p < 0.001 | An increase in user trust is almost exclusively associated with alerts that are perceived as precise. | 323 |
| Alert arrival wrt shaking | 186.29 | 4 | p < 0.001 | Receiving an alert before the shaking is the single most powerful factor in building user trust. | 258 |

| AEA vs others | 47.5 | 6 | p < 0.001 | An alert that beats other sources of information is a confidence booster. | 54 |
| Alert mode | 15.64 | 2 | p < 0.001 | A silent notification is perceived as a system failure that leads to a significant loss of trust. | 60 |

Our data demonstrate a user's confidence in the system escalates when the alert is perceived as precise, and conversely, diminishes when deemed inaccurate. A shift in trust is independent of the alert's accuracy. The perception of an alert's accuracy is profoundly influenced by its clarity and timeliness. Users generally deem alerts received during or after an event to be "inaccurate" because their delayed nature makes them misleading. Difficulty in comprehending information significantly increases the likelihood of its perception as inaccurate. Clear communication is therefore essential to build trust and ensure the credibility of the information shared (Table S1). Most users recalled information about the earthquake's estimated magnitude (27.7%) and epicentral distance (23%). In contrast, only 7.4% of posts mentioned the safety measures provided in AEA alerts (Figure S1).

The utility of an alert serves as the primary determinant of user trust; specifically, an alert that provides actionable preparation time is paramount in fostering user confidence. Prior to ground shaking, an alert's arrival bolsters user trust, whereas alerts received during or after shaking have the opposite effect, significantly reducing a system's credibility. Furthermore, trust in the AEA system is intrinsically linked to its timeliness relative to other EEW systems. User confidence is significantly boosted by early alerts (with respect to the other EEW alerts), while late alerts have the opposite effect. The perceived precision of an alert (for instance the magnitude estimate), however, has much less impact in these situations. Our data also shows that the alert's mode significantly affects a user's future trust. An early audible alert effectively builds and maintains trust, whereas some users perceive a silent notification as a system failure, resulting in a considerable loss of confidence.

# Discussion

Our understanding of how humans respond to EEW systems is still developing, often based on anecdotal evidence, making a comprehensive analysis of their effects difficult (Cochran and Husker, 2019; Bossu et al., 2021; Fallou et al., 2022; Vaiciulyte et al., 2022; Saunders and Wald 2025; McBride et al., 2019, 2022). This study analyzes the user's perception of performance of Google's AEA system during the Mw 6.2 Marmara Ereğlisi, Türkiye earthquake. Social media posts from the X platform turned out to be particularly insightful, offering valuable perspectives on user interactions with EEW alerts and their behavior during the earthquake. However, it is important to clarify that because these findings are derived from cross-tabulations and chi-square tests of perceptual reports, the observed statistical relationships strictly reflect users' subjective perceptions and associative behaviors, rather than causal evidence for the engineering accuracy or systemic latency of the AEA system itself. The integration of LLMs with unconventional data, such as social media posts, can furnish EEW operators and researchers with crucial user feedback regarding perceptions and expectations of specific alerts. This capacity can then inform social scientists who study human behavioral responses to natural disasters and warning information, which facilitates the evaluation of disaster reduction programs.

This earthquake, though not causing significant damage, occurred in close proximity to Istanbul, a densely populated metropolitan area within a country characterized by high seismic activity and a recent history of catastrophic seismic events, such as the 2023 Mw 7.8 and Mw 7.5 Kahramanmaraş Earthquake Sequence. This geographical and historical context inherently elevates public sensitivity to seismic hazards and, consequently, to the perceived efficacy of mitigation tools like EEW systems. The responses of alert recipients during such critical periods represent a pivotal measure of an EEW system's real-world utility. Furthermore, the analysis of X post data pertaining to this earthquake offers invaluable insights into the behavioral responses of an earthquake-experienced population within a seismically active nation to earthquake alerts.

However, it is important to acknowledge the inherent limitations of this approach. Specifically, the demographic assessment in this study relies on limited and indirectly inferred information. Systematic data regarding key demographic variables—such as user age, education level, and socioeconomic status—are not available through this passive collection method. Under these circumstances, it remains unclear to what extent the observed user behaviors, emotional responses, or perceptions of the system can be meaningfully associated with specific demographic characteristics.

Furthermore, the findings derived from this study highlight several actionable recommendations and prospective areas for future research:

- *Optimizing Alert Design:* The significant relationship between alert sound, informational clarity, user response, perceived utility, and trust highlights the critical role of effective alert design. Therefore, future efforts should concentrate on: I) Investigating multimodal alert designs that integrate auditory cues with clear, concise visual information, such as dynamic text displays, simple infographics, or brief animated instructions, to reinforce protective actions; II) Exploring diverse alert sounds or rhythmic patterns engineered to universally capture attention without inducing undue panic, ensuring they effectively override "do not disturb" settings and are distinguishable from other phone notifications; and III) Continuing to refine the language and presentation of safety information within alerts to maximize comprehension, particularly for individuals experiencing heightened fear or surprise.

- *Further Behavioral Research*: Unstructured social media data provided valuable insights, but a comprehensive understanding of user behavior requires additional targeted research such as: I) Further research is needed to understand why specific emotional states, such as fear, surprise, and anxiety, did not strongly correlate with particular actions; II) Sustained research is needed to monitor how user trust and perceived system utility evolve across various earthquake experiences and diverse alert outcomes, including false alarms, timely warnings, and delayed alerts; III) Analyze the influence of comprehensive demographic variables (beyond inferred gender, incorporating age, education, and socioeconomic status) and socio-cultural contexts on alert interpretation and behavioral responses, as well as the impact of online communication styles (e.g., sarcasm, humor) on LLM emotion classification, likely requiring the integration of structured user surveys; and IV) Investigating the ideal psychological thresholds for varying warning durations and their impact on the likelihood and characteristics of protective actions taken.

# Conclusion

This study provides an analysis of the performance of Google's Android Earthquake Alert (AEA) system during the Mw 6.2 Marmara Ereğlisi, Türkiye earthquake on April 23, 2025. By leveraging LLMs to analyze 511 public social media posts from the X platform, we extracted 42 distinct attributes related to user experience and behavior, offering insights into the real-world utility and perception of smartphone-based EEW systems.

Our findings actionable insights for optimizing alert design, emphasizing the need for multimodal alerts with clear, concise visual and auditory information. It also underscores the importance of public education campaigns to shape user expectations, clarify appropriate behavioral responses, and foster trust by demonstrating the tangible benefits of early warning. Future research should delve deeper into the interplay of emotional states and actions, the evolution of user trust over time, the influence of demographic and socio-cultural factors, and the psychological thresholds for varying warning durations.

## Acknowledgements

We would like to thank the reviewers for insightful comments. An anonymized summary dataset of the extracted 42 attributes from X posts is available at https://github.com/smousavi05/eew-xposts. All figures in this paper were generated using Python version 3.6.

# Supplementary Materials for

## Early Warnings of Millions of Users for 2025 Mw 6.2 Marmara Ereğlisi, Türkiye Earthquake by Android's Smart-Phone-Based Alert System

Corresponding author. Email: mousavim@google.com

**The PDF file includes:**
    Figs. S1 to S2
    Tables S1 to S2

TABLE S1 Other Significant Relationships Among Other Variables.

| Variable | Chi-square Statistic | DF | Significance | Relationship | Total observations (N) |
|---|---|---|---|---|---|
| AEA Info Accuracy vs Helpfulness Reason | 280.36 | 2 | $p < 0.001$ | For users, an alert's accuracy and its helpfulness are essentially the same thing. An alert is only perceived as precise if it is useful (i.e., provides time to prepare or confirms the event). If the alert is not useful because it arrived too late, it is overwhelmingly written off as inaccurate, regardless of its technical correctness. | 332 |
| Alert Arrival WRT Shaking vs AEA Info Accuracy | 172.2 | 2 | $p < 0.001$ | An alert's perceived accuracy is fundamentally defined by its arrival time. An alert is only seen as "precise" if it arrives before the shaking begins, providing a useful warning. Alerts that arrive during or after the event are overwhelmingly perceived as "inaccurate," as their failure to be timely renders them incorrect in the user's mind. | 275 |
| Users Sentiment vs Alert Arrival WRT Shaking | 147.69 | 8 | $p < 0.001$ | A user's sentiment is fundamentally tied to the alert's arrival time. An alert that arrives before the shaking is a successful intervention that fosters positive feelings. An alert that arrives during or after the shaking is a failure that generates negative sentiment. | 317 |
| Helpfulness Reason vs Post Alert Action | 83.63 | 4 | p < 0.001 | People who valued the alert for giving them time to prepare were far more likely to take an Active Response.<br>The perceived usefulness of an alert is strongly tied to the subsequent action. When users feel an alert gives them time to prepare, they overwhelmingly choose to take active measures. When the alert serves as confirmation, they are more likely to become passively aware. | 97 |
| Alert Info Clearance vs Helpfulness Reason | 46.78 | 2 | p < 0.001 | When an alert is not timely, users are also far more likely to perceive its content as confusing or unclear. | 144 |
| Users Sentiment vs AEA vs Other EEW | 43.08 | 9 | p < 0.001 | A user's sentiment is fundamentally tied to the alert's timeliness. An alert that arrives earlier than other sources generates positive feelings. An alert that arrives later generates negative feelings. Speed is the primary determinant of a positive user experience. | 58 |
| Alert Arrival WRT Shaking vs Alert Info Clearance | 15.97 | 2 | p < 0.001 | An alert's perceived clarity is strongly dependent on its arrival time. Alerts received during the shaking are significantly more likely to be found unclear, as the chaotic environment hinders comprehension. Alerts that provide a warning before the shaking are more likely to be understood clearly. | 118 |
| AEA Info Accuracy vs Alert Info Clearance | 31.36 | 1 | p < 0.001 | A user's perception of an alert's accuracy is heavily dependent on its clarity. If users cannot easily understand the information, they are highly likely to perceive it as inaccurate. Clear communication is essential for building trust and ensuring the information is perceived as credible. | 149 |
| Felt Shaking vs Reason for Taking No Action | 21.08 | 4 | p < 0.001 | The reason for inaction is strongly tied to whether the user physically felt the shaking. When the shaking is not felt, inaction is often a deliberate choice because it is deemed unnecessary. When the shaking is felt, inaction is more likely a reaction to the event itself, caused by a lack of | 88 |

| | | | | time or confusion. | |
|---|---|---|---|---|---|
| Alert Mode vs Users Sentiment | 19.96 | 4 | p < 0.001 | A user's sentiment is strongly dependent on the alert's mode. A silent notification is highly likely to generate negative sentiment. A positive or neutral sentiment is almost exclusively associated with receiving an audible alert. | 77 |
| Alert Type vs Helpfulness Reason | 11.88 | 2 | p < 0.001 | The perceived helpfulness of an alert is directly tied to its type. "Take Action" alerts are valued because they provide time to prepare. "Be Aware" alerts are valued because they confirm the event. Each alert type is successfully fulfilling its intended purpose in the user's mind. | 89 |

Table S2: The prompt used for the analysis of X posts.

```
You are a computational social scientist analyzing the user's experiences
about the performance of Google's Earthquake Early Warning system (a.k.a.
Android Earthquake Alert or AEA) during recent earthquakes in Türkiye.
Phones plugged in and stationary report their availability for earthquake
monitoring by AEA. An on-phone detection algorithm analyzes acceleration
time series for sudden changes indicative of seismic P- or S-wave arrivals.
Upon detecting a potential event, de-identified parameter data is sent to
the backend server. This detection capability is deployed as part of Google
Play Services, core system software, meaning it is on by default for the
vast majority of Android smartphones and does not require activation or
installation of any additional application. The servers then match the
pattern of phone triggering with possible seismic sources in the time-space
domain. An earthquake is declared and its source parameters (e.g.,
magnitude, hypocenter, and origin time) are estimated. Upon detection of an
earthquake, the intensity of the ground shaking and its potential extent are
estimated. For events with estimated magnitude exceeding M4.5, AEA sends two
distinct types of alerts to users that are within the impacted area. These
include "Take Action" and "Be Aware" alerts for users within the regions
expected to experience moderate or greater (i.e., >= MMI 5) and weak (i.e.,
MMI 3 or 4), respectively. The "Take Action" alert takes over the entire
screen of the phone, breaking through any do-not-disturb settings and makes
a characteristic sound designed to be attention grabbing. The "Be Aware"
alert appears as a notification similar to other phone or app notifications,
but with a characteristic sound. Once the shaking has passed, or if the
alert arrives after shaking, the alerts are replaced by the "Earthquake
Occurred" notification. The delivered alerts to the Android phone users
contain a short summary of the event attributes, precautionary instructions,
earthquake safety info, and a short user survey feedback of the alert
delivery. Please use the provided examples as reference and extract these
information from the input tweet if available: Username; Post's date-time in
YYYY-MM-DDTHH:MM format, keeping in mind that screenshots of tweets are
taken in EST while local time would be in Istanbul time; Geolocation (like
the city or town if available); How many seconds before the earthquake did
they receive the alert?; Does the post include a screenshot or picture of
```

```
the received alert (YES, NO)?; What is the time of the issued alert --  in
YYYY-MM-DDTHH:MM format -- shown on the attached image of the received alert
to the post?; What is the magnitude of the earthquake shown on the attached
image of the received alert to the post?; What is the distance (in miles) to
the earthquake shown on the attached image of the received alert to the
post?; What is the language of received alert?; Does the post include the
alert contour and user's relative position (YES, NO, NOT_APPLICABLE)?; What
is the approximate location of the user (shown by blue circle marker on the
alert notification)?; What type of the alert they receive
(BE_AWARE_NOTIFICATION, TAKE_ACTION_ALERT, UNKNOWN)?; What is the alert
source (AEA, EQN, ETC)?; What is the overall sentiment of the replies to the
post (CONFIRMATION_OF_POSITIVE_POST, CONFIRMATION_OF_NEGATIVE_POST,
OPPOSITION_OF_POSITIVE_POST, OPPOSITION_OF_NEGATIVE_POST, NOT_APPLICABLE)?;
Did the user feel the earthquake shaking (YES, NO, UNKNOWN)?; Did the alert
come with a sound notification or was it just a text notification
(ALERT_WITH_SOUND, SILENT_NOTIFICATION, UNKNOWN)?; What action did the
person take after receiving the alert (DROP_COVER_HOLD_ON, EVACUATED,
MOVED_TO_SAFETY, PROTECTED_OTHERS, PASSIVE_AWARE, SOUGHT_INFO,
WARNED_CONTACTED_OTHERS, NO_ACTION)?; What is the sentiment of the user
(POSITIVE, NEGATIVE, NEUTRAL, MIXED)?; Beyond general sentiment, did the
user express specific emotions regarding the alert or the earthquake (FEAR,
ANXIETY, REASSURANCE, GRATITUDE, CONFUSION, SURPRISE, ANNOYANCE)?; How
helpful or unhelpful was the earthquake alert from the user's point of view
(NOT_HELPFUL, HELPFUL, VERY_HELPFUL, NEUTRAL)?; Did the user think the
system could be improved (YES, NO, UNKNOWN)?; When did the alert arrive
(BEFORE_SHAKING, DURING_SHAKING, AFTER_SHAKING, UNKNOWN)?; What was the
level of the shaking that the user felt (STRONG, WEAK, UNKNOWN)?; What was
the intensity of the ground shaking in MMI scale at the user's location
based on post content (1, 2, 3, 4, 5, 6, 7, 8)?; Did someone else near the
user receive an earthquake alert as well (YES, NO, UNKNOWN)?; Where was the
user when received the alert (INDOOR, OUTDOOR, UNKNOWN)?; Was the user alone
or with others when receiving the alert (YES, NO, UNKNOWN)?; Was it the
first time the user received an alert from an earthquake alerting system
(YES, NO, UNKNOWN)?; What was their past experience receiving an earthquake
alert (POSITIVE, NEGATIVE, NEUTRAL, UNKNOWN)?; Did the user experience
earthquake damage in the past (YES, NO, UNKNOWN)?; What was the user's
gender (FEMALE, MALE, LIKELY_FEMALE, LIKELY_MALE, UNKNOWN)?; What specific
information from the alert did the user recall or mention (SAFETY_ADVICE,
ESTIMATED_MAGNITUDE, ESTIMATED_DISTANCE,
ESTIMATED_INTENSITY_AT_THEIR_LOCATION, ALERT_SOURCE like 'Android Earthquake
Alerts System)?; Did the user comment on the accuracy of the information
provided by the AEA (was the magnitude, location, or timing perceived as
correct or incorrect when compared to their experience or other sources,
PRECISE, INACCURATE)?; Did the user mention any technical issues with
receiving or viewing the alert itself (POWER_LOSS, ALERT_SCREEN_FREEZING,
ALERT_SOUND_ISSUE, ALERT_NOT_APPEARING_WHEN_EXPECTED, UNKNOWN)?; Did the
user comment on how clear or easy to understand the alert message and any
instructions were (CLEAR_TO_UNDERSTAND, ALMOST_CLEAR_TO_UNDERSTAND, UNCLEAR,
"UNKNOWN")?; If the user stated they took no specific protective action
after receiving the alert did they provide a reason why (NO_TIME, CONFUSION,
DEEMED_UNNECESSARY, OTHER_REASON_FOR_NO_ACTION, REASON_UNSPECIFIED)?; Did
the user's post suggest a level of trust (or distrust) in the AEA system for
future earthquake events based on this particular experience
(WILL_TRUST_MORE, WILL_TRUST_LESS, NEUTRAL, UNKNOWN)?; If the user
explicitly stated the alert was helpful or unhelpful, what specific reasons
did they give for this assessment (PROVIDED_TIME_TO_PREPARE,
```

CONFIRMED_IT_WAS_AN_EARTHQUAKE, IT_ARRIVED_TOO_LATE_TO_BE_USEFUL)?; Did the user compare the Android Earthquake Alert to any other earthquake warning systems they might be aware of or other sources of earthquake information (AEA_ALERT_ARRIVED_EARLIER, AEA_ALERT_ARRIVED_LATER, AEA_ALERT_WAS_MORE_PRECISE, AEA_ALERT_WAS_LESS_PRECISE, UNKNOWN)? If no information is provided for a specific key, use the tag "UNKNOWN" for that key. Make sure to limit your response to a JSON format containing only the following keys: "username", "post_datetime", "post_location", "warning_time_seconds", "with_alert_screenshot", "alert_time", "magnitude_on_alert_screenshot", "distance_on_alert_screenshot_ml", "alert_language", "alert_screenshot_with_contour", "user_approximate_location_on_alert", "alert_type", "alert_source", "reply_sentiment", "felt_shaking", "alert_mode", "post_alert_action", "users_sentiment", "users_emotion", "helpfulness", "system_improvement", "alert_arrival_wrt_shaking", "shaking_level", "shaking_intensity_mmi", "alert_received_by_others", "indoor_vs_outdoor", "user's_accompany", "first_earthquake_alert_experience", "user's_past_earthquake_experience", "past_earthquake_damage_experience", "user's_gender", "alert_info_recall", "aea_info_accuracy", "technical_issues_with_alert", "alert_info_clearance", "reason_for_taking_no_action", "future_trust_level", "helpfulness_reason", "aea_vs_others", "reasoning". Let's walk this through step by step with sample data. Sample data: { "screenshot_1": { "username": "@cinnamonjemur", "post_datetime": "2025-04-23T12:59", "post_location": "West of Istanbul", "warning_time_seconds": "UNKNOWN", "with_alert_screenshot": "YES", "alert_time": "2025-04-23T12:49", "magnitude_on_alert_screenshot": "5.3", "distance_on_alert_screenshot_ml": "88.0", "alert_language": "English", "alert_screenshot_with_contour": "YES", "user_approximate_location_on_alert": "West of Istanbul, near the Marmara Sea coast", "alert_type": "BE_AWARE_NOTIFICATION", "alert_source": "AEA", "reply_sentiment": "CONFIRMATION_OF_POSITIVE_POST", "felt_shaking": "YES", "alert_mode": "UNKNOWN", "post_alert_action": "UNKNOWN", "users_sentiment": "POSITIVE", "users_emotion": "GRATITUDE", "helpfulness": "VERY_HELPFUL", "system_improvement": "NO", "alert_arrival_wrt_shaking": "UNKNOWN", "shaking_level": "UNKNOWN", "shaking_intensity_mmi": "UNKNOWN", "alert_received_by_others": "UNKNOWN", "indoor_vs_outdoor": "UNKNOWN", "user's_accompany": "UNKNOWN", "first_earthquake_alert_experience": "UNKNOWN", "user's_past_earthquake_experience": "UNKNOWN", "past_earthquake_damage_experience": "UNKNOWN", "user's_gender": "LIKELY_MALE", "alert_info_recall": [ "ESTIMATED_MAGNITUDE", "ESTIMATED_DISTANCE", "ALERT_SOURCE" ], "aea_info_accuracy": "PRECISE", "technical_issues_with_alert": "UNKNOWN", "alert_info_clearance": "CLEAR_TO_UNDERSTAND", "reason_for_taking_no_action": "UNKNOWN", "future_trust_level": "WILL_TRUST_MORE", "helpfulness_reason": "CONFIRMED_IT_WAS_AN_EARTHQUAKE", "aea_vs_others": "UNKNOWN", "reasoning": "The user expresses positive sentiment ('çok iyi' - 'very good') and posts a screenshot of the AEA alert. The date-time of the post was converted from EST to Istanbul time (EST+7). The alert screenshot shows key details like magnitude (5.3) and distance (88.0 miles). The replies confirm the user's positive experience, with another user asking how to enable it and the original poster replying that it works automatically. The user's positive feedback and the nature of the information shared suggest they found the alert helpful and accurate, thereby increasing their trust in the system. Many fields are marked 'UNKNOWN' as the user's short tweet does not provide details on their actions, emotions, or specific experience of the shaking." }, "screenshot_2": { "username": "@yigitech", "post_datetime": "2025-04-24T07:30", "post_location": "Marmara Region", "warning_time_seconds": "21",

"with_alert_screenshot": "YES", "alert_time": "UNKNOWN",
"magnitude_on_alert_screenshot": "4.6", "distance_on_alert_screenshot_ml":
"41.6", "alert_language": "Turkish", "alert_screenshot_with_contour": "NO",
"user_approximate_location_on_alert": "NOT_APPLICABLE", "alert_type":
"BE_AWARE_NOTIFICATION", "alert_source": "AEA", "reply_sentiment":
"NOT_APPLICABLE", "felt_shaking": "YES", "alert_mode": "UNKNOWN",
"post_alert_action": "UNKNOWN", "users_sentiment": "POSITIVE",
"users_emotion": "UNKNOWN", "helpfulness": "VERY_HELPFUL",
"system_improvement": "NO", "alert_arrival_wrt_shaking": "BEFORE_SHAKING",
"shaking_level": "UNKNOWN", "shaking_intensity_mmi": "UNKNOWN",
"alert_received_by_others": "UNKNOWN", "indoor_vs_outdoor": "UNKNOWN",
"user's_accompany": "UNKNOWN", "first_earthquake_alert_experience":
"UNKNOWN", "user's_past_earthquake_experience": "UNKNOWN",
"past_earthquake_damage_experience": "UNKNOWN", "user's_gender": "MALE",
"alert_info_recall": [ "ESTIMATED_MAGNITUDE", "ALERT_SOURCE" ],
"aea_info_accuracy": "PRECISE", "technical_issues_with_alert": "UNKNOWN",
"alert_info_clearance": "UNKNOWN", "reason_for_taking_no_action": "UNKNOWN",
"future_trust_level": "WILL_TRUST_MORE", "helpfulness_reason":
"PROVIDED_TIME_TO_PREPARE", "aea_vs_others": "AEA_ALERT_ARRIVED_EARLIER",
"reasoning": "The user directly compares the performance of Google's Android
alert system (AEA) with another application, 'Deprem Ağı' (Earthquake
Network). The tweet explicitly states that the Android alert arrived 21
seconds before the earthquake, while the other app's alert arrived 15
seconds before, making AEA faster by 6 seconds. This constitutes a positive
sentiment and a direct reason for the alert's helpfulness
('PROVIDED_TIME_TO_PREPARE'). The user provides a screenshot from the
'Deprem Ağı' app, not the AEA alert itself, which is why fields like
'alert_time' are unknown. The post time is estimated based on the earthquake
time mentioned in the screenshot plus the 11 minutes mentioned in the tweet
text. The user's name 'Yiğit' is male. The distance was converted from km to
miles (67km ≈ 41.6 miles)." } } Extract similar information from the
following tweet:

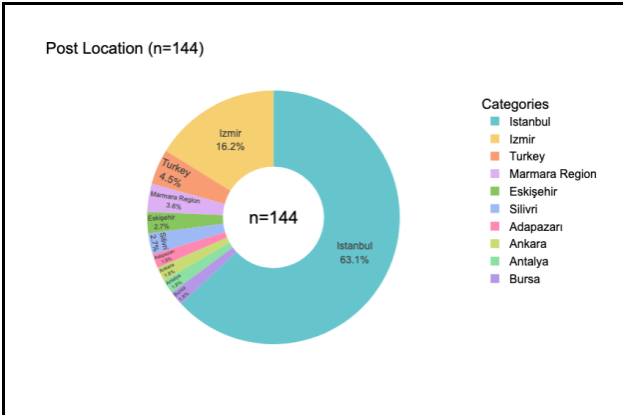
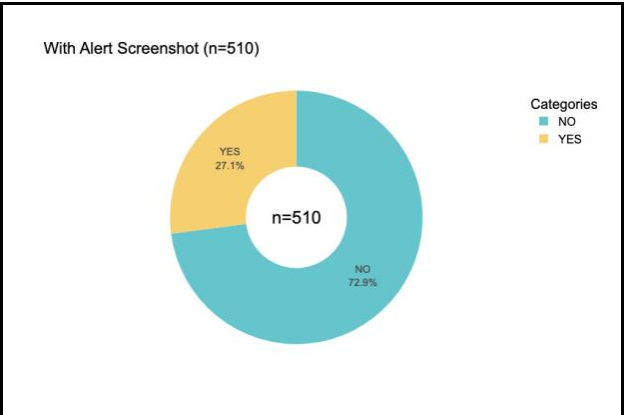
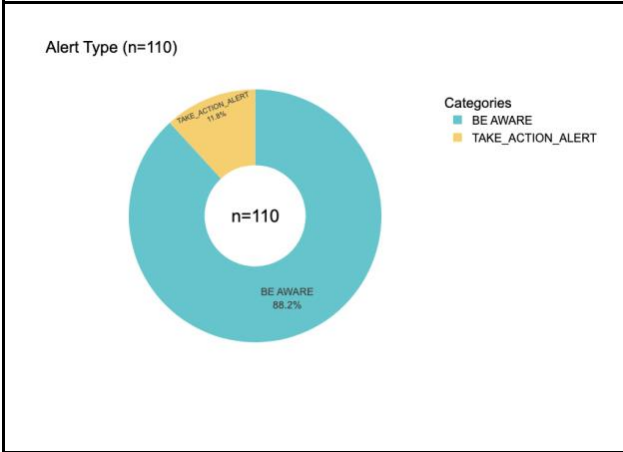
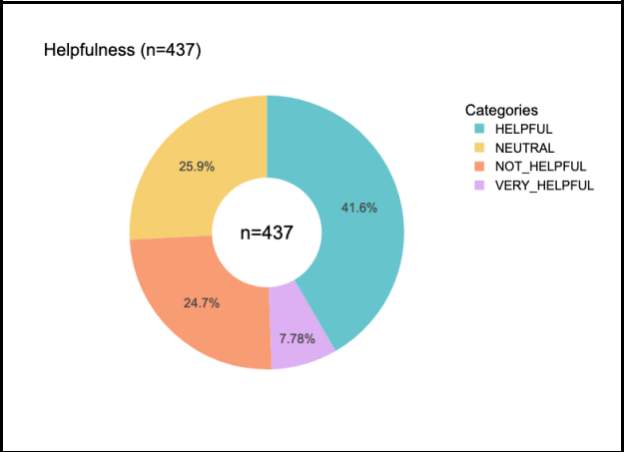
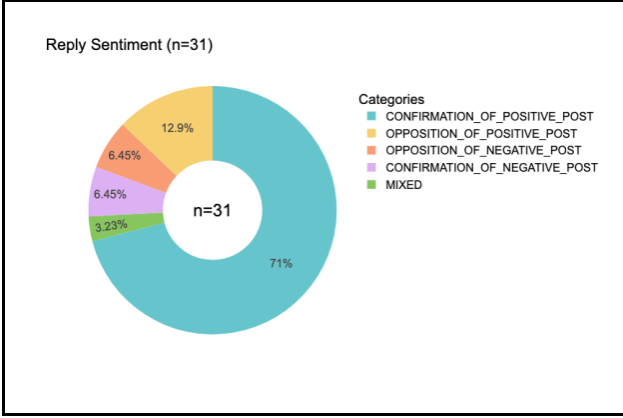
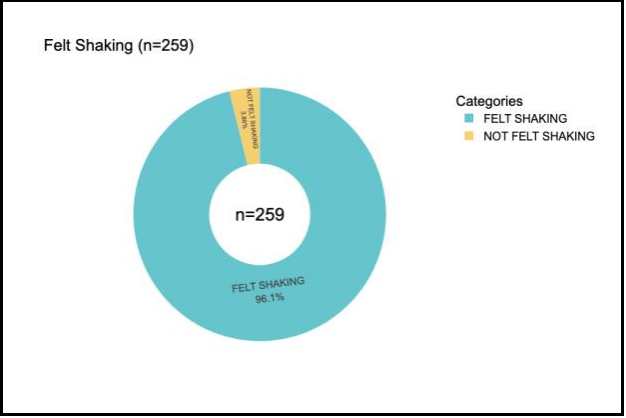

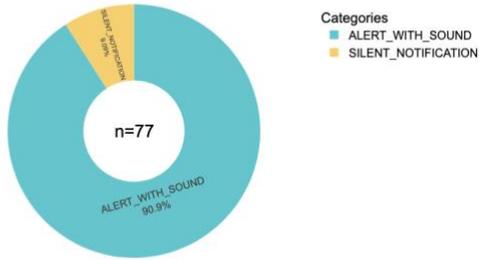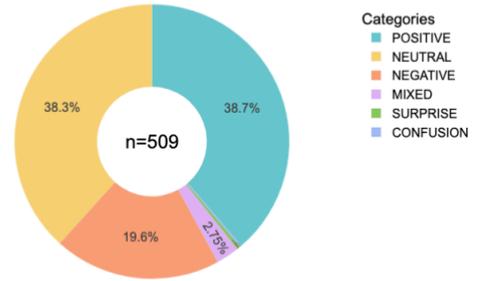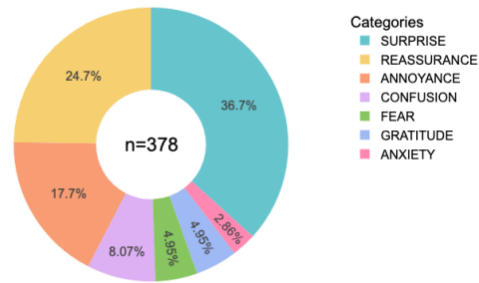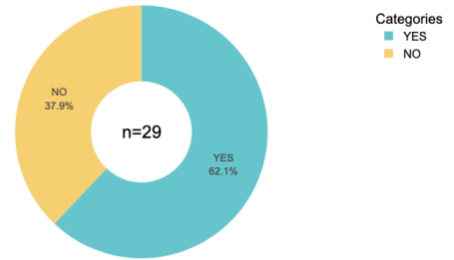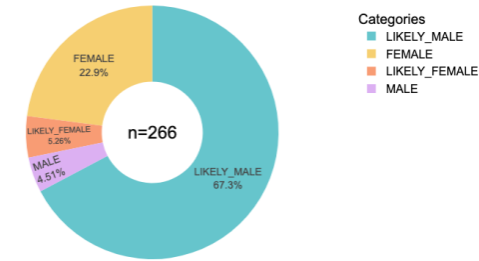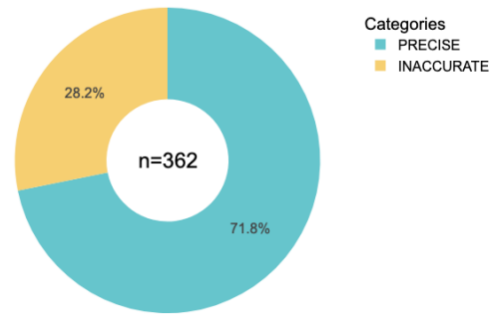

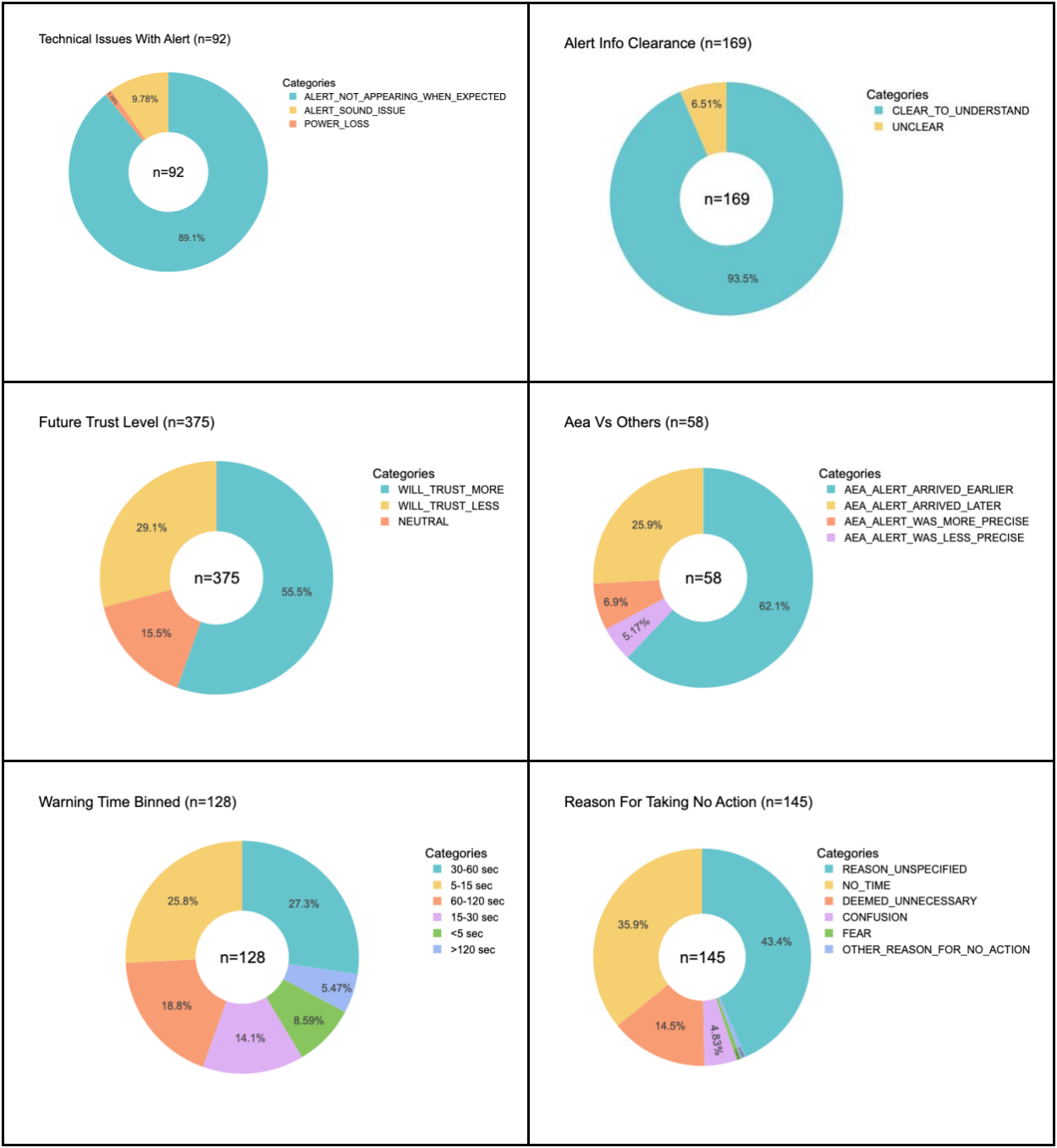

Figure S1: Distributions of extracted information from X posts for various attributes.

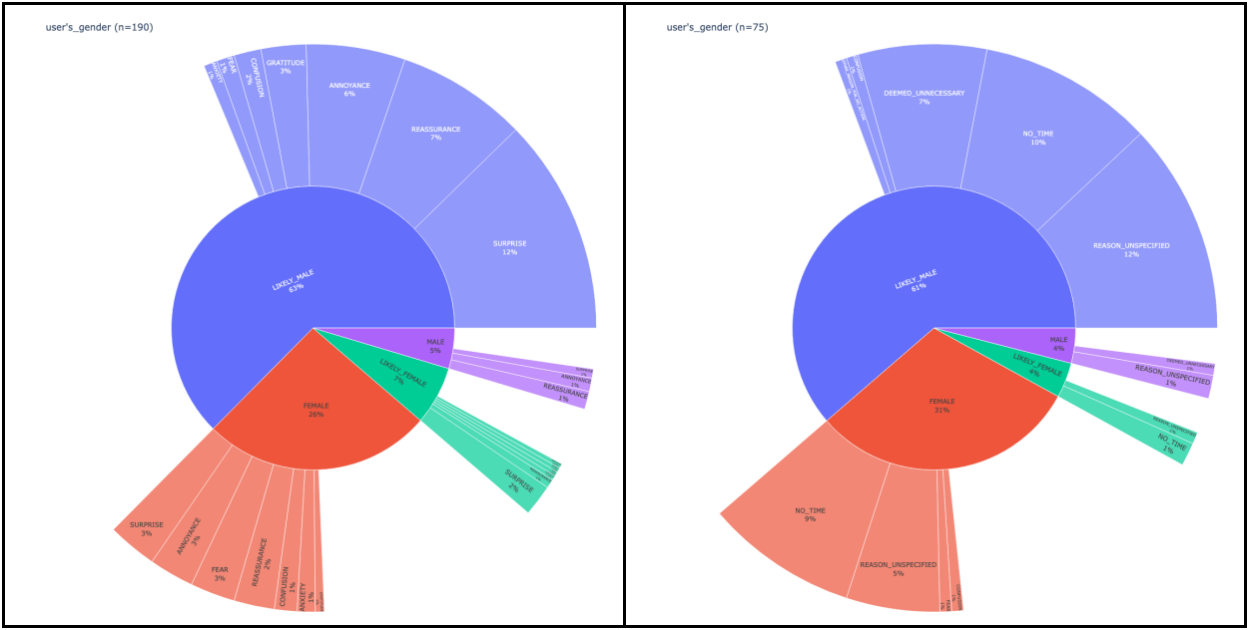

Figure S2: Demographic distributions of different attributes (extracted information from X posts) with respect to the user gender.